\documentclass[conference,a4paper]{IEEEtran}
\usepackage{cite}
\usepackage{epsfig,graphics,mathrsfs,amssymb,amsmath,bm,color,yfonts,txfonts,multirow,multicol,slashbox,dblfloatfix}
\usepackage{subfigure}
\usepackage{algorithm2e}
\usepackage[noend]{algpseudocode}

\begin{document}

\title{Precoded Cluster Hopping in Multi-Beam High Throughput Satellite Systems}
\author{Mirza Golam Kibria, Eva Lagunas, Nicola Maturo, Danilo Spano and Symeon Chatzinotas\\
SIGCOM Research Group, SnT, University of Luxembourg\\
}
\maketitle

\begin{abstract}

Beam-Hopping (BH) and precoding are two trending technologies for the satellite community. While BH enables flexibility to adapt the offered capacity to the heterogeneous demand, precoding aims at boosting the spectral efficiency. In this paper, we consider a high throughput satellite (HTS) system that employs BH in conjunction with precoding. In particular, we propose the concept of Cluster-Hopping (CH) that seamlessly combines the BH and precoding paradigms and utilize their individual competencies. The cluster is defined as a set of adjacent beams that are simultaneously illuminated. In addition, we propose an efficient time-space illumination pattern design, where we determine the set of clusters that can be illuminated simultaneously at each hopping event along with the illumination duration. We model the CH time-space illumination pattern design as an integer programming problem which can be efficiently solved. Supporting results based on numerical simulations are provided which validate the effectiveness of the proposed CH concept and time-space illumination pattern design.

\end{abstract}

\IEEEpeerreviewmaketitle
\section{Introduction}

The next-generation high throughput satellite systems (HTS) deliver a significant enhancement over available capacity. The satellite resources are very scarce and extremely expensive. Furthermore, satellites are powered by solar energy and/or battery. The lower the payload mass/size, the smaller the energy consumption. Therefore, these resources need to be utilized efficiently. Compared to the conventional global beam satellite system, HTS brings about new technical challenges in terms of efficient utilization of the satellite resources such as spectrum, transponders/spot-beams, and power. The satellite traffic is not homogeneous over time and coverage areas. The satellite traffic exhibits non-uniform patterns and multiple time zones in demand distribution. Offering the same level of service or capacity over the whole coverage area all the times results in inefficient utlization of available resources. The heavily loaded beams will suffer from capacity deficiency while lightly loaded beams will have the excessive capacity. The necessity of supporting uncertain and uneven traffic demand over time and space throughout the satellite coverage area has been studied and recognized in various activities including several European Space Agency (ESA) projects \cite{11,12,ESA}. The satellite must have flexibility in the allocation of the resources to beams in order to optimize the scarce and expensive satellite resources.

It may not be able to power all the beams simultaneously. Furthermore, the satellite payload mass is a very strong constraint as the satellite launching cost increases with the payload mass. Studies have shown that a limited number of RF chains under satellite power constraint can serve a large number of beams by virtue of beam-hopping (BH). BH provides the means to flexibly adapt the offered capacity to the time and geographic variations of the traffic demands\cite{3}. In a BH system, at any given time only a subset of the satellite beams are illuminated. The set of illuminated beams changes in each hopping event based on a time-space transmission pattern\cite{4,6,9}. Therefore, with BH, all the available satellite resources are employed to provide service to a certain subset of beams. One of the main benefits of BH is the resource allocation flexibility in the time domain by allocating different dwell time (duration over which a particular beam is illuminated) to different beams depending on the demand. Another major advantage of employing BH is the reduction on the payload mass as not all the beams are activated at the same time. Although the BH solution promises some major advantages mentioned above, it brings in some technical/operational challenges, for example, designing an illumination pattern able to perfectly match the demands, acquisition, and synchronization of bursty transmitted data, the exploitation of extra degrees of freedom provided by the fact that certain regions of the coverage area are inactive.

While standard technologies, e.g., the DVB-S2\cite{DVBS2}, embrace interference avoidance approach that employs frequency-reuse of the available spectrum amongst beams, more recent paradigms endorse different approach through the management and the exploitation of the interference amongst beams with the aim of enhancing the spectral efficiency. The objective is to maximize the use of the user link available spectrum, which represents a limited resource of the satellite system. One of the major challenges in such full frequency-reuse (FFR) systems is dealing with intra-beam and inter-beam interference. Intra-beam interference can be easily avoided by scheduling only one user in a beam at a time. Multi-user multiple input single output digital signal processing techniques, such as linear precoding, can be applied in the forward link of a multibeam satellite system operating in FFR for spatial interference mitigation\cite{Vazquez,Zheng,Joroughi}. Moreover, with the help of precoding, the system flexibility can be improved through the efficient management of satellite resources, e.g., on-board power. At the same time, since all beams operate at the same spectral band, it allows any user terminal to jointly receive power from adjacent beams, thus creating a beam-free paradigm and offering an additional degree of flexibility which consists in exploiting the received power that is intended to the neighboring beams.

Clearly, BH provides the means to flexibly adapt the offered capacity to the spatio-temporal variations of the traffic demands, while precoding exploits the multiplexing feature enabled by the use of multiple antenna feeds at the transmitter side onboard to boost the spectral efficiency. These two effective strategies can create unique opportunities if they are properly combined. Focusing on the convergence of both BH and precoding techniques, we propose a cluster hopping (CH) approach in which a set of clusters are illuminated simultaneously. Note that the simultaneously illuminated clusters can be adjacent or non-adjacent. In the case of non-adjacent clusters, the system will suffer from inter-cluster interference. We impose the constraint that the simultaneously illuminated clusters are non-adjacent to avoid inter-cluster interference. The CH procedure allows higher frequency reuse by placing inactive clusters as barriers for co-channel inter-cluster interference. We propose an efficient solution to the problem of designing the time-space transmission plan, therefore, finding the optimal sets of clusters to be illuminated at different hopping events along with their corresponding illumination periods under payload size (number of power amplifiers), payload architecture and satellite power constraints. The proposed CH scheme seamlessly combine the BH and precoding paradigms and utilize their individual competencies. 

The proposed cluster hopping offers flexibility from two different perspectives: i) long-term flexibility: aiming at the management and assignment of satellite resources so as to match the offered capacity with the average beam demands, and  ii) short-term flexibility: once the illumination pattern is fixed, the optimization of the user scheduling within the clusters can further improve the overall traffic matching at a user-level in a beam-free paradigm. Supporting simulation results are provided which validate the effectiveness of the proposed scheme.

\section{System Model and CH Illumination Pattern Optimization Problem Statement}
Let us consider a high throughput multi-beam satellite system with $N_{\rm B}$ beams. The $N_{\rm B}$ beams are divided into $N_{\rm C}$ clusters. The satellite system employs FFR, i.e., all the beams operate over the whole satellite spectrum $B_{\rm W}$ Hz. In addition, we assume that the satellite system operates on two orthogonal, i.e., vertical and horizontal polarizations. 

In this study, we focus on the forward link (gateway-satellite-user) of a broadband multi-beam satellite system.

 In this paper, we use the following terminologies:
\begin{itemize}
\item {\it Cluster}: A group of adjacent beams illuminated simultaneously. The available satellite beams are grouped into a number of clusters. A given beam may belong to only one cluster or multiple clusters. Clusters can be of any shape, e.g., compact shaped and non-compact shape. Compact clusters increase the distance between two simultaneously illuminated clusters, thus, makes the clusters less susceptible to inter-cluster interference.
\item {\it Snap-shot}: A particular arrangement of illuminated and non-illuminated clusters. There can be as many as $2^{N_{\rm C}}$ achievable snap-shots. Note that not all the snap-shots are valid in a sense that only a given number of non-adjacent clusters can be illuminated simultaneously because of payload limitations on the total amplified bandwidth as well as not to have inter-cluster interference. Therefore, in a valid snap-shot, a given number of non-adjacent clusters (i.e., the beams within those clusters) are illuminated, while the remaining clusters are inactive.
\item {\it Slot-time}: Duration of a time-slot. The slot-time defines the time granularity of the hopping operation, i.e., slot-time is the minimum illumination period for a selected snap-shot.
\item {\it Hopping Window}: Consists of a number of time-slots. The service time (hopping operation time) is divided into hopping windows. The duration of the hopping window represents the repetition period of the BH satellite system.
\item {\it Illumination Ratio}: Ratio of number of simultaneously illuminated beams to total number of beams in the system during a given time-slot or time instant.
\end{itemize}
\begin{figure}
  \centering
   \includegraphics[scale=.42]{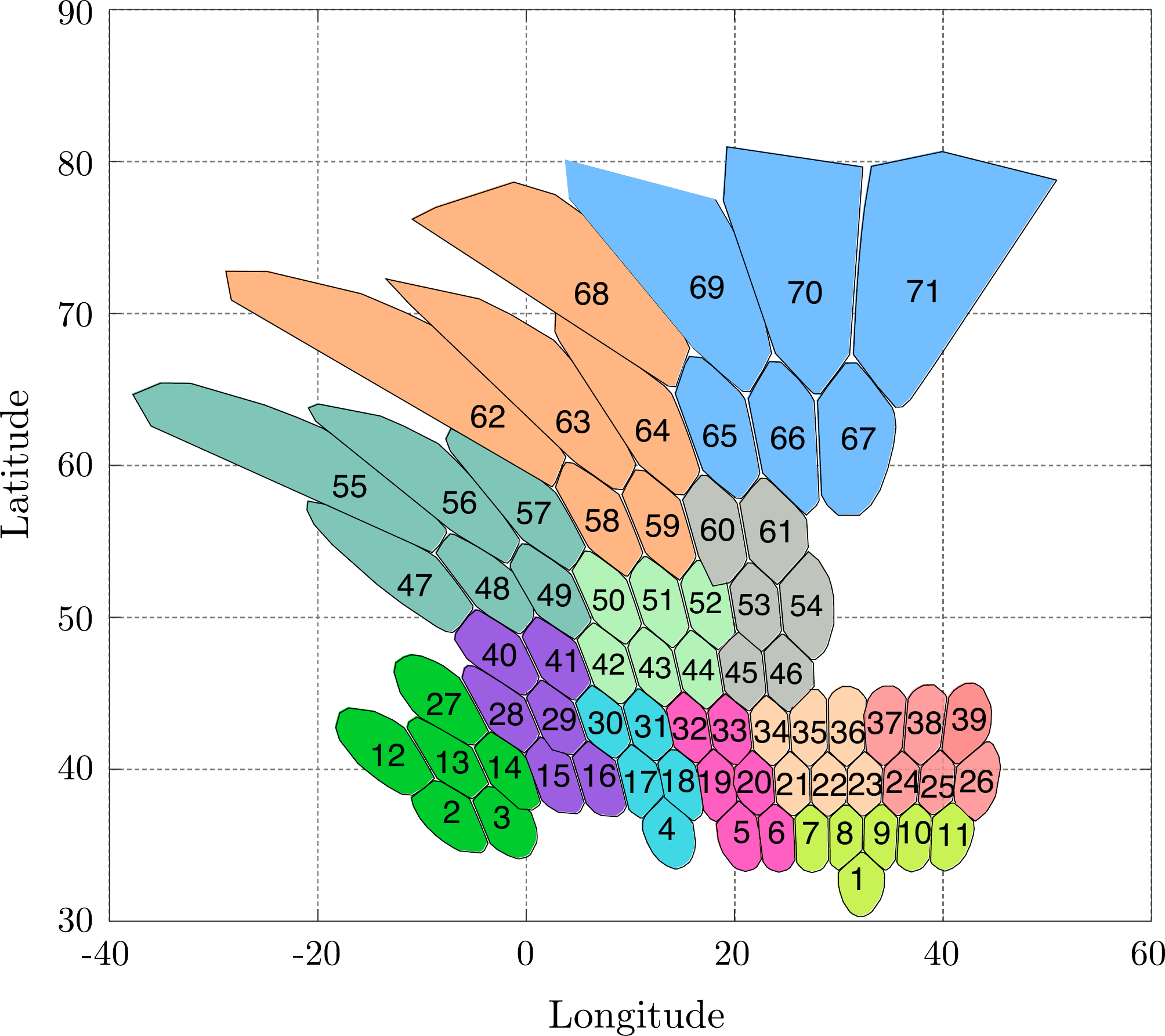}
   \caption{Clustering of beams from 71 beam-pattern. Beams with same color form a cluster. There are 12 clusters. All the clusters have 6 beams except one with 5 beams.}
   \label{fig_FPD_1}
\end{figure}

In our considered system model, each beam $i$ has an associated demanded capacity $\hat{d}_i$ (measured in bps), which is known. We arrange all the beam demanded capacity into a single vector 
$\hat{\bf d}\in \mathbb{R}^{N_{\rm B\times 1} }\triangleq[\hat {d}_1,\hat {d}_2,\cdots,\hat {d}_{N_{\rm B}}]^T$. Here, $(\cdot)^T$ stands for transpose operation.
Let the vector $\bf d$ stores the demands (in bps) from all the clusters in the multibeam satellite system as
${\bf d}\in \mathbb{R}^{ N_{\rm C}\times1 }\triangleq[d_1,d_2,\cdots,d_{N_{\rm C}}]^T,$
where $d_j$ is the aggregated demand from all the beams in cluster $j$ calculated as
$$d_j=\sum_{\forall i}\hat {d}_i,i\in \mathcal{A}_j,$$
where $\mathcal{A}_j$ is the set of indices of the beams in cluster $j$.
The hopping window is set to $T_{\rm H}$, which is equally divided into $N_{\rm slot}$ time-slots. So, we have $T_{\rm H}=N_{\rm slot}\times T_{\rm slot}$. 
We opt to design the appropriate illumination pattern, i.e. the sets of clusters to be illuminated over different hopping instants within the hopping cycle $T_{\rm H}$ and the duration of the hopping events in order to offer an average capacity  as close as possible to the requested demand ${\bf{d}}$. Average values are of interest as the active users will change over different snap-shots.

\subsection{CH Illumination Pattern Optimization Problem}
The objective is to find snap-shots at different hopping events such that the demands of the users are fairly satisfied. Let ${\bf c}=[c_1,c_2,\cdots,c_{N_{\rm C}}]^T$ define the cluster capacity vector of size $N_{\rm C}\times 1$, where $c_j$ is the capacity of cluster $j$. Since the optimization, i.e., selection of the most suitable snap-shots along with their illumination periods is performed for each individual hopping window, we scale down the cluster demand and supply capacity as $m_j=T_{\rm H} d_j,\hspace{1mm} (j=1,2,\cdots,N_{\rm C})$ and $p_j=T_{\rm slot} c_j,\hspace{1mm}  (j=1,2,\cdots,N_{\rm C})$, respectively. Note that ${\bf m}=[m_1,m_2,\cdots,m_{N_{\rm C}}]^T$ and ${\bf p}=[p_1,p_2,\cdots,p_{N_{\rm C}}]^T$ are measured in bits per $T_{\rm H}$  and bits per $T_{\rm slot}$, respectively. Let the supply vector be given by ${\bf s}=[s_1,s_2,\cdots,s_{N_{\rm C}}]^T$, where $ s_j$ is the cluster supply capacity in response to demand $m_j$. The CH illumination pattern design problem  is formulated to maximize the minimum ratio between the offered capacity and the requested demand among the {cluster/beams} and can be expressed as follows
\begin{equation}
\label{main123x}
 \begin{array}{*{35}{l}}
\hspace{10.5mm}\underset{{\bf u}_1,{\bf u}_2,\cdots, {\bf u}_{N_{\rm slot}}}{\max} \underset{i}{\min}\hspace{2mm}\frac{s_j}{m_j}\\
\text{}\text{subject to }\hspace{2mm}\sum_{j=1}^{N_{\rm C}}u_{n,j}=N_{\rm P}, n=1,2,\cdots, N_{\rm slot} \\
\text{}\hspace{15mm} \hspace{2.8mm}{\bf u}_n^T{\bf Au}_n=0,\hspace{1mm}, n=1,2,\cdots, N_{\rm slot} \\
\text{}\hspace{15mm} \hspace{2.8mm}{\bf s}=\sum_{n=1}^{N_{\rm slot}}\left({\bf u}_n\odot {\bf p}\right)
\end{array}
\end{equation}
The vectors ${\bf u}_1,{\bf u}_2,\cdots, {\bf u}_{N_{\rm slot}}$ are the optimization variables, where ${\bf u}_n$ is a binary vector of size $N_{\rm C}\times 1$, and $u_{n,j}\in\{0,1\}$ is the $j$th element in ${\bf u}_n$ corresponding to cluster $j$. The positions of 1's in ${\bf u}_n$ give the indices of the illuminated clusters during a given time-slot. 
The second constraint enforces that active clusters are not adjacent to each other by means of matrix ${\bf A}\in \{0,1\}^{N_{\rm C}\times N_{\rm C}}$, which is the binary adjacency matrix of the clusters. It is a square symmetric matrix, i.e., ${\bf A}_{i,j}={\bf A}_{j,i}$. If ${\bf A}_{i,j}=1$, the cluster $i$ is adjacent to cluster $j$. $N_{\rm P}$ is the number of illuminated clusters in each time-slot. Here $\odot$ denotes element-wise multiplication. Note that the constraint $\sum_{j=1}^{N_{\rm C}}u_{n,j}=N_{\rm P},\forall n$ restricts the system to illuminate a set of $N_{\rm P}$ clusters in each time-slot, and constraint ${\bf u}_n^T{\bf Au}_n=0$ confirms the non-adjacency of the simultaneously illuminated clusters. The problem in \eqref{main123x} is a binary quadratic problem and it is very difficult to solve. We propose an efficient and linear formulation for the optimization problem.

\section{Precoding for Mitigating Intra-Cluster Interference}
Since every beam is served using the same frequency resource (FFR among beams), co-channel interference should be mitigated or addressed. We employ linear precoding to minimize the interference. Therefore, the achievable capacity of user $k$ in beam $i$ (of cluster $j$) is strongly linked to channel gains of all scheduled users in other beams of the cluster $j$ and the precoding design.

For simplicity, in the following, we assume that a single user is served by an active beam. In principle, there can be as many users as the number of beams per cluster distributed over the cluster coverage.  By assuming FFR among the beams in a cluster, the collection of received signals at the beam-center users is modeled as
\begin{equation}
\label{eqn-1}
{\bf y}_j={\bf H}_j{\bf W}_j{\bf x}_j+{\bf z}_j,
\end{equation}
where ${\bf H}_j\in\mathbb{C}^{\Pi_j\times \Pi_j}$ is the channel matrix pertaining to cluster $j$ and $\Pi_j$ denotes the cardinality (number of cells/beams) of cluster $j$.  ${\bf y}_j\in\mathbb{C}^{\Pi_j\times1}$  is the vector of received signals while  ${\bf x}_j\in\mathbb{C}^{\Pi_j\times1}$ is the vector unprecoded data symbols destined for beam-center users in cluster $j$. ${\bf z}_j$ is the collection of noise power at the corresponding beam-center users of cluster $j$, and ${\bf W}_j$ is the linear precoder employed in cluster $j$.

We employ minimum mean square error (MMSE) precoding\cite{Peel}. The MMSE precoding matrix for cluster $j$ is given by

\begin{equation}
\label{eqn-4}
{\bf W}_j=\beta^j_{\rm MMSE}{\bf H}_j^H\left({\bf H}_j{\bf H}_j^H+\frac{1}{P}{\bf \tau}{\bf I}_{\Pi_j}\right)^{-1}.
\end{equation}
Here, ${\bf I}_{\Pi_j}$ is the square identity matrix with number of row/columns equal to $\Pi_j$ and ${\bf \tau}=[\tau_{j,i_1},\tau_{j,i_2},\cdots,\tau_{j,i_{\Pi_j}}]$ is $1\times \Pi_j$ vector of noise powers in the beams of cluster $j$. $P=P_{\rm T}/N_{\rm B}$ is the transmit power per beam, where $P_{\rm T}$ is th total power available at the satellite. Therefore, there is no power sharing among the feeds since we consider that the payload is not equipped with multiport amplifier. Unlike zero-forcing precoding that cancels the intra-cluster interfrence completely, MMSE enhances the SNIR of the desired signal reducing the overall error variance. In addition, a scaling factor $\beta^j_{\rm MMSE}$ is introduced in order to meet the per beam transmit power consraint $\left\{{\bf W}_j{{\bf W}_j}^H\right\}_{i,i}\le P_{\rm T}/N_{\rm B}$ and can be determined as $\beta^j_{\rm MMSE}=\frac{P_{\rm T}/N_{\rm B}}{\max\left(\text{diag}\left({{\bf W}_j}^H{\bf W}_j\right)\right)}$
\begin{figure*}
  \centering
   \includegraphics[scale=.45]{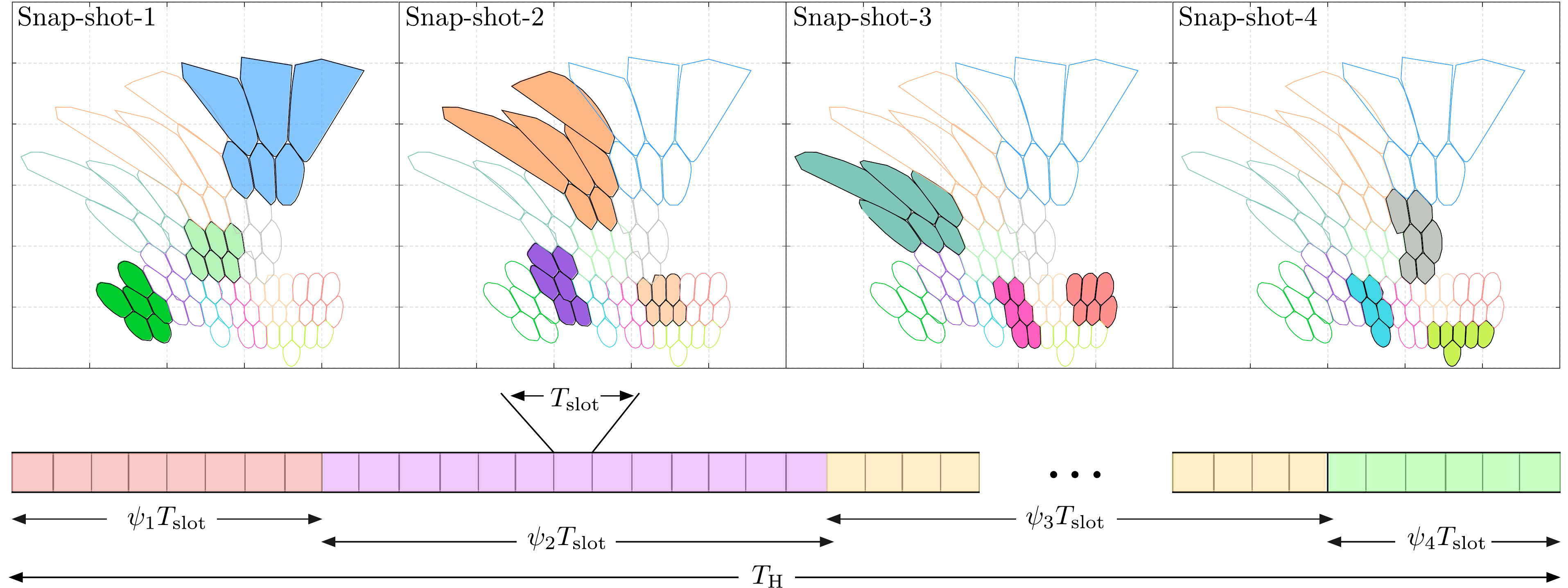}
   \caption{An example of hopping pattern and illumination period optimization scenario. In this particular case, 4 optimal snap-shots are selected. The illumication periods are also optimized so as to satisfy the demands of the beam-center users.}
   \label{fig_FPD_02}
\end{figure*}
Now, the received SNIR for the $k$th beam-center user in cluster $j$ can be expressed as 
\begin{equation}
\label{eqn-XX}
\text{SNIR}_k=\frac{\left |\left({\bf H}_j{\bf W}_j\right)_{k,k}\right |^2}{\sum_{i\neq k}^{\Pi_j}\left |\left({\bf H}_j{\bf W}_j\right)_{k,i}\right |^2+\tau_{j,k}}.
\end{equation}
Using the SNIR-spectral efficiency mapping according to DVB-S2, we perform a mapping of the SNIR in \eqref{eqn-XX} and calculate the capacity of each beam. The cluster capacities are collected in ${\bf c}$, where $c_j$ is the summation of capacities of all the beams in cluster $j$ according to
$c_j=\sum_{\forall i}r_i, i\in \mathcal{A}_j,$
where $r_i$ is the capacity of beam $i$. Note that the individual beam capacity is calculated by taking into account the polarization multiplexing, i.e., capacity of a cluster is the summation of capacities achievable in vertical and horizontal polarization. As mentioned earlier, we obtain $p_j$ as $T_{\rm slot}c_j$.

\section{Proposed Solution For CH Illumination Pattern Optimization}
In this section, we propose an efficient solution for the optimization problem in \eqref{main123x}. Note that optimization problem in \eqref{main123x} is a combinatorial problem whose search space is huge and makes the optimization computationally expensive. 
To address this, we propose to use Graph Theory to limit the research spaces by discarding the solutions that do not satisfy the constraint ${\bf u}_n^T{\bf Au}_n=0$. Next, to refine the search space, we additionally discard those solutions that do not satisfy the $N_{\rm P}$ constraint, i.e., only $N_{\rm P}$ non-adjacent clusters can be active simultaneously.
The procedure is detailed in the following.

\subsection{Generation of Valid Snap-shots}
We create the list of pairwise non-adjacent clusters, where each set of pairwise non-adjacent cluster is referred to as ``independent set". An independent set is defined as the set of clusters in which no two clusters are adjacent. Let ${\bf G}\in \{0,1\}^{N_{\rm C}\times Q}$ be the binary matrix that contains all the snap-shots with exactly $N_{\rm P}$ active clusters (corresponds to the positions of 1's in each column vector) and the clusters can be adjacent or non-adjacent. Let  $\binom{N_{\rm C}}{N_{\rm P}}$ returns a matrix $\bf Z$ (of size $N_{\rm P}\times Q$) containing binomial coefficient or all, i.e., $Q$ possible combinations of the elements $\{1,2,\cdots, N_{\rm C}\}$ taken $N_{\rm P}$ elements at a time. The positions of 1's in each column of $\bf G$ is defined by the elements of the correponding column (one-to-one mapping) in $\bf Z$. For example, let $N_{\rm P}=3$ and the 3rd column vector of $\bf Z$ has the elements $\{3\hspace{1mm} 5\hspace{1mm} 6\}$, then the 3rd column of $\bf G$ will have 1's in position 3, 5 and 6.
 Since the adjacent clusters are not allowed to be illuminated simultaneously, we need to filter out the snap-shots with adjacent clusters. We, therefore, apply the following test to check whether the set of clusters in each snap-shot (in ${\bf Z}$, i.e., in $\bf G$) are non-adjacent or not. The clusters in $i$-th column of $\bf G$, ${\bf g}_i$ yields an independent set, i.e., a valid snap-shot, only if ${\bf  g}_i^T{\bf Ag}_i=0$ holds.
Accordingly, we find all the independent sets or sets of non-adjacent clusters with a given cardinality $N_{\rm P}$ and store them in ${\bf V}^{N_{\rm C}\times N_{\rm ss}}$, where $N_{\rm ss}$ is the number of valid snap-shots. Each column of $\bf V$, ${\bf v}_i$ corresponds to a valid snap-shot and has exactly $N_{\rm P}$ 1's.

Once we have generated the valid snap-shot matrix $\bf V$ and the cluster capacity vector $\bf p$ is available, we can perform the hopping pattern and illumination period optimization. Note that generation of valid snap-shots inherently deals with first two constraints in \eqref{main123x}. Consequently, the hopping pattern and illumination period optimization can now be expressed as
\begin{equation}
\label{main123xx}
 \begin{array}{*{35}{l}}
\hspace{10.5mm}\underset{{\bf u}_1,{\bf u}_2,\cdots, {\bf u}_{N_{\rm slot}}}{\max} \underset{i}{\min}\hspace{2mm}\frac{s_j}{m_j}\\
\text{}\text{subject to }\hspace{2mm}{\bf s}=\sum_{n=1}^{N_{\rm slot}}\left({\bf u}_n\odot {\bf p}\right) \\
\end{array}
\end{equation}
with ${\bf u}_n\in\left\{{\bf v}_1,{\bf v}_2,\cdots,{\bf v}_{N_{\rm ss}}\right\}$. A typical hopping pattern and illumination period optimization scenario is depicted in Fig.~\ref{fig_FPD_02}. Here, the number of hopping events is 4. The 4 optimal snap-shots are chosen along with their optimal illumination periods so as to fairly satisfy the demands of the satellite users.

\subsection{Proposed Solution}
We opt to design the appropriate illumination pattern, i.e. the sets of clusters to be illuminated over different hopping instants within the hopping cycle $T_{\rm H}$ and the duration of the hopping events in order to offer an average supply capacity  $s_j$ as fairly (among the clusters) and as closely as possible to the requested demand $m_j$. From the cluster-beam mapping in Fig.~\ref{fig_FPD_1} and the valid snap-shot matrix ${\bf V}$, we generate a matrix ${\bf L}\in\mathbb{R}^{N_{\rm C}\times N_{\rm ss}}$ where $n$th column vector in ${\bf L}$ is given by ${\bf l}_n={\bf v}_n\odot {\bf p} $. So, ${\bf L}$ contains all possible illumination patterns together with corresponding supply capacity per $T_{\rm slot}$.

Note the the satellite payload power constraint is infused in the problem in terms of the number of clusters simultaneously illuminated. We can simplify the $\max-\min$ optimization problem by turning it into a maximization problem with the help of an additional slack variable $t$ along with a new constraint $\frac{s_j}{m_j}\ge t\triangleq s_j\ge m_jt$. 
We can model this as an integer quadratic program. Let $ \psi_i\in\mathbb{Z}$ ( $\mathbb{Z}$ is a plane/space of nonnegative integer numbers) be a nonnegative integer variable indicating the illumination duration for snap-shot $ i$ $(i=1,2,\cdots,N_{\rm ss})$ in terms of the number of time-slots. Now, we can rewrite ${\bf{s}}$ as
\begin{equation} 
{\bf s}=\sum_{i=1}^{N_{\rm ss}}\left({\bf{l}}_{i}\psi_i\right).
\end{equation}
Now, adding one more constraint to maintain the defined hopping window as written below
\begin{equation}
\sum_{i=1}^{N_{\rm ss}}\psi_i=N_{\rm slot} \text{ or }\sum_{i=1}^{N_{\rm ss}}\psi_iT_{\rm slot}=T_{\rm H},
\end{equation}
the problem of finding the best cluster-hopping pattern in \eqref{main123x} can be equivalently expressed as
\begin{equation}
\label{main123}
 \begin{array}{*{35}{l}}
\hspace{9.5mm}\underset{\psi_i,i=1,2,N_{\rm ss}}{\max} \hspace{1mm}t\\
\text{}\text{subject to} \hspace{3mm}{\bf{s}}=\sum_{i=1}^{N_{\rm ss}}\left({\bf l}_i\psi_i\right),\vspace{1.5mm} \\
\text{}\hspace{17mm}{\bf{s}}\succeq t{\bf{m}},\\
\text{}\hspace{16mm}\sum_{i=1}^{N_{\rm ss}}\psi_i=N_{\rm slot}, \vspace{1.5mm} \\
\end{array}
\end{equation}
Here, the relation ${\bf x}\succeq {\bf y}$ states that an element in $\bf x$ succeds the same indexed element in $\bf y$, i.e., $x_i\ge y_i$. The optimization problem formulation in \eqref{main123} can be solved in a much easier way than \eqref{main123x}, since it is an integer linear programming problem, for which computationally efficient solutions are known. A distinct advantage is the possibility to obtain the optimal illumination period for each snap-shot. The number of hopping events (i.e., the number of optimal snap-shots) is given by the number of non-zero $\psi_i$ and the corresponding illumination period is given by the values of non-zero $\psi_i$. 
The optimization problem in \eqref{main123} is an integer programming problem and can be efficiently solved using integer programming solver\cite{CVX}.

 \begin{table}[h]
\centering
\caption{Simulation Parameters} 
\label{tab-123}
\begin{tabular}{l |r}
\hline
\hline
Satellite longitude & \hspace{10mm}$13^{\circ}{\rm E}$ (GEO)\\ 
Satellite totalradiated power, $P_{\rm T}$ &\hspace{2mm}6000 W\\
Number of HPA, $N_{\rm HPA}$ & \hspace{2mm}36 (2 beams per HPA)\\
Number of beams, $N_{\rm B}$ & \hspace{2mm}71\\
Beam radiation pattern & \hspace{2mm}Provided by ESA\\ 
Downlink carrier frequency & \hspace{2mm}19.5 GHz\\ 
Number of clusters, $N_{\rm C}$ & \hspace{2mm}12\\ 
User link bandwidth, $B_{\rm W}$ & \hspace{1.95mm}500 MHz \\ 
Roll-off factor & \hspace{1.95mm}20\% \\ 
Duration of a time-slot, $T_{\rm slot}$ & \hspace{2mm}1.3 ms\\
Hopping Window,  $T_{\rm H}$ & \hspace{1.95mm}256 $T_{\rm slot}$ \\ 
Illumination Ratio &  approx. 1/4\\
\hline
\end{tabular}
\end{table}

\begin{figure}
  \centering
   \includegraphics[scale=.6]{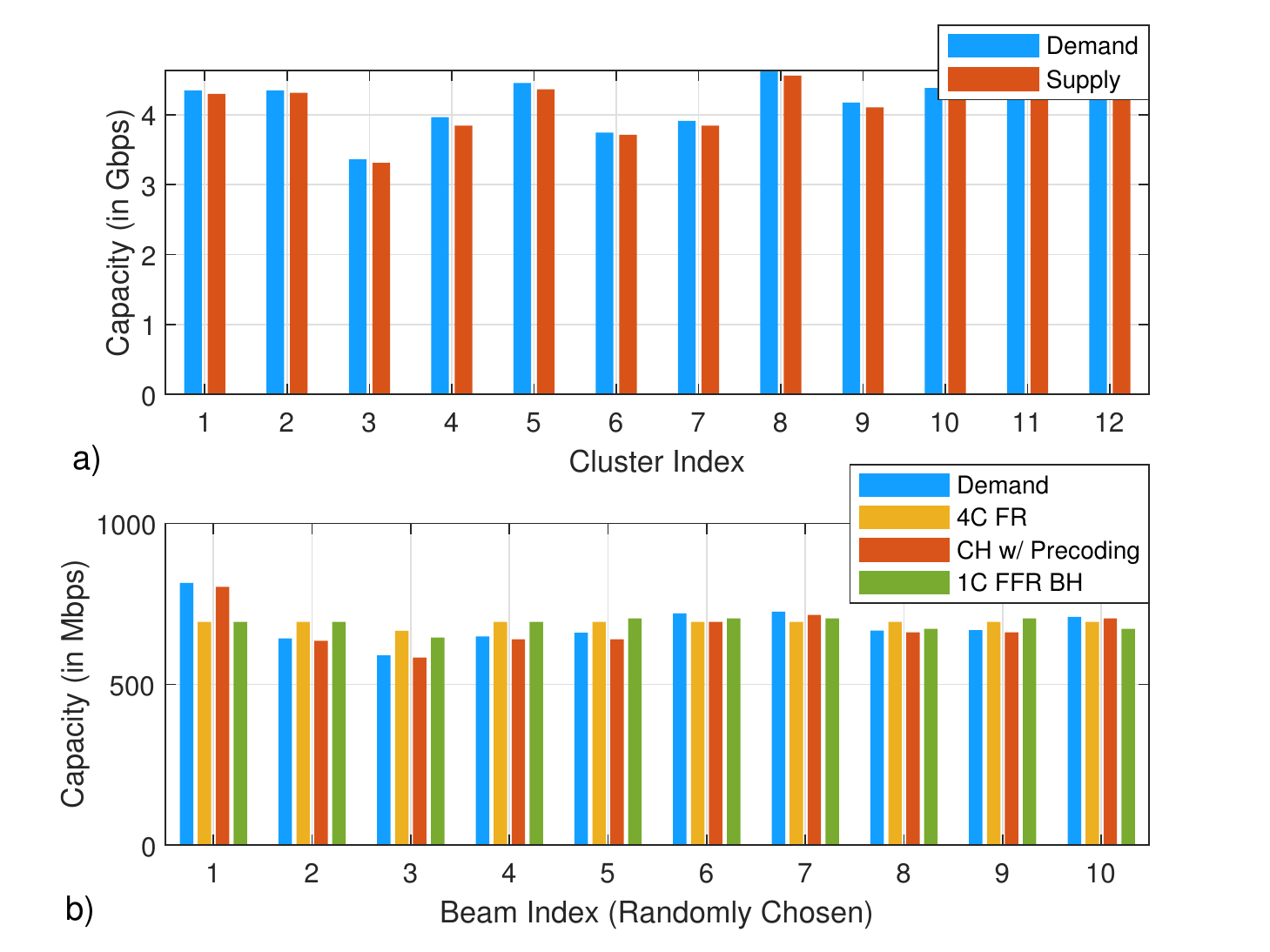}
   \caption{a) Demand capacity vs. offered capacity at cluster level, and b) demand vs. offered capacity at beam level (for a set of randomly selected users).}
   \label{fig_flex_1}
\end{figure}

\begin{figure}
  \centering
   \includegraphics[scale=.44]{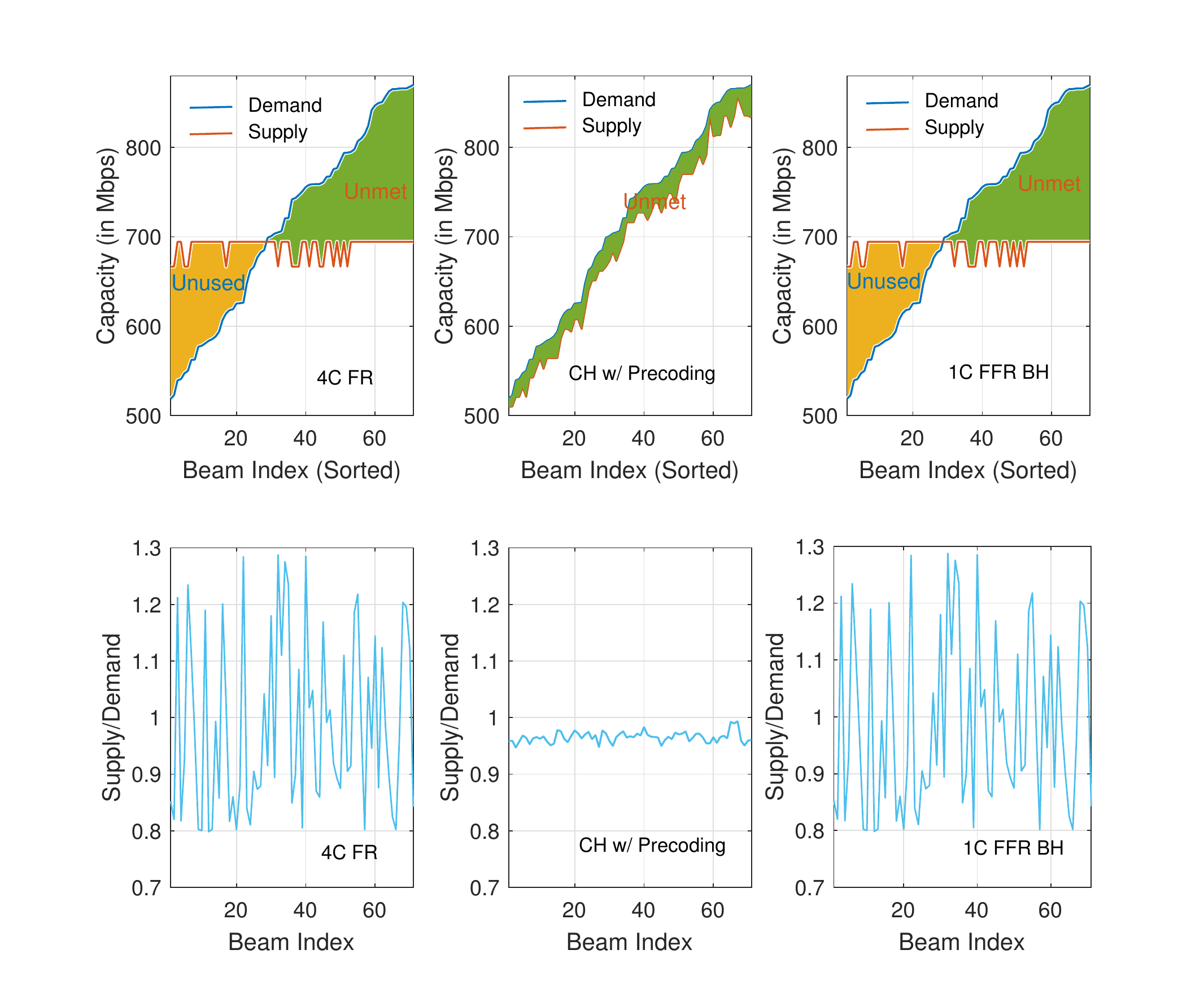}
   \caption{a) Unmet vs. unused capacity. b) Fairness among the beams in terms of the ratios of supply capacity to demand capacity. }
   \label{fig_flex_2}
\end{figure}

\section{Simulation Results}

In this section, we evaluate the proposed CH snap-shot selection and illumination period optimization results. The simulation parameters are listed in Table.\ref{tab-123}. In our simulation, for CH scheme, both horizontal and vertical polarities dwell on the same beam simultaneously. The performance of the proposed time-division based CH scheme is compared to that of frequency division method, i.e., 4-color frequency reuse (4C FR) scheme with both vertical and horizontal polarization and 1-color FFR BH (1C FFR BH) scheme. The perfromance of the CH scheme has been compared with the performances of the abovementioned benchmark schemes under the same conditions by system configurations.

Fig.~\ref{fig_flex_1}a shows the performances of different resource allocation schemes in terms of cluster demand and offered capacity. We can see that the cluster-hopped (with precoding) satellite payload exhibits very good rate matching capability. Both the benchmark solutions work at beam level. In order to do the comparison between the proposed CH scheme and the benchmark solutions, we assume that the cluster offered capacity can be distributed across the beams as we have a beam-free paradigm in CH, which is achieved with proper user scheduling within a cluster not included in this paper for space limitations. In particular, in the current evaluation, we redistribute the cluster capacity among the beams proportionally based on their demands. Notice that the CH scheme performs exceptionally well compared to the 4C FR scheme as the 4C FR cannot adapt the supply capacity to the demand, as seen in Fig.~\ref{fig_flex_1}b. For the 4C FR and 1C FFR BH schemes, the offered capacity remains almost unchanged, no matter what the beam demands are. Therefore, for non-uniform traffic distribution (over time/space), the the 4C FR and 1C FFR BH (with current setup) schemes are highly inefficient. In case of 1C FFR BH, there are strict constraints in the possible illumination patterns due to interference avoidance.

Note that in 1C FFR BH, the illumination ratio is approximately $1/4$. Unfortunately, the $1/4$ approach significantly limits the demand matching capabilities since one beam can be a hot beam in a certain coverage area while another may be a cold beam in another coverage area. If the active beam ratio is reduced (e.g., from $1/4$ to $1/8$), the flexibility can be improved at the expenses of longer BH cycle so as to ensure that all beams are illuminated at a certain time. However, assuming a fix coverage area, the only way to reduce the illumination ratio in 1C FFR BH or conventional BH is to modify the antenna array by adding more elements, i.e., more beams. With CH, this is not required, as we can vary the cluster size. In the proposed CH, the illumination ratio does not apply, as we can deal with interference. Therefore, CH has much more flexibility by properly designing the size of the clusters.

In Fig.~\ref{fig_flex_2}a, we evaluate the performance of the proposed hopping solution and the other schemes in terms of their capabilities in reducing the unmet and unused capacities. In the $x$-axis, the beams are sorted according to their demand. It is clear from the performance curves that the proposed CH scheme outperforms other schemes in minimizing the unused as well as unmet capacity as it has the ability to adapt the snap-shot selection and illumination period to the demand. However, the actual performance will depend on the performance of the user scheduling method. For the 4C FR scheme, the area between the curves before the cross-over/intersection defines the unused capacity while the area between the curves after the cross-over point quantifies the unmet capacity. Both 4C FR and the 1C FFR BH scheme exhibit similar tendencies. While on the other hand, the proposed CH scheme has either the unmet capacity of unused capacity because of the nature of the optimization (based on fairness) we have adopted. According to our observations, the proposed CH scheme outperforms the benchmark solutions in most of the cases. The unmet and unused capacity for any scheme depends on the demands of the clusters. In Fig.~\ref{fig_flex_2}b, we investigate the capability of the proposed CH approach in incorporating fairness as well as delivering proportional capacity among the clusters and compare with other solutions. It is clear that the proposed CH solution provides very good match between the demand and offered capacity.

\section{Conclusions}

The study in this paper aims at assessing the benefits of introducing CH in multi-beam satellite system. The non-uniform and clustered demand distribution over the coverage are the basic reasons to introduce CH with precoding. 
We have found that CH scheme can efficienctly accommodate the variation or non-uniformity in the traffic distribution by means of selecting the most suitable snap-shots and illuminating each snap-shot proportional to the traffic demand. 

In this study we have limited our CH scheme to have clusters with approximately equal number of beams. Demand based dynamic clustering can have some impact on the performance, where clusters have varying number of beams. The analysis of such flexible clustering based CH scheme is left as future works. In addition, user-scheduling will mainly define how beam demands are actually being treated. The combined impact of proposed time-space transmission plan and user-scheduling on overall system performance will also be investigated in our future works.
The performance of proposed CH solution with lower illumination ratio needs to be compared to BH solution with only non-adjacent beams being activated simultaneously with the same illumination ratio in order to evaluate the efficacy of precoded cluster hopping.

\section*{Acknowledgement}
This work has received funding from the European Space Agency (ESA) funded activity FlexPreDem: Demostrator of Precoding Techniques for Flexible Broadband Systems. The views of the authors of this paper do not necessarily reflect the views of ESA.

\end{document}